\documentstyle[amssymb,twoside,fleqn,espcrc2]{article}

\input{tcilatex}

\begin{document}

\author{Remo Garattini\thanks{%
Talk given at the {\it Third Meeting on Constrained Dynamics and Quantum
Gravity, }Villasimius (Sardinia), September 13-17, 1999.} \\
Facolt\`{a} di Ingegneria, Universit\`{a} degli Studi di Bergamo,\\
Viale Marconi, 5, 24044 Dalmine (Bergamo) Italy.\\
M\'{e}canique et Gravitation, Universit\'{e} de Mons-Hainaut,\\
Facult\'e des Sciences, 15 Avenue Maistriau, \\
B-7000 Mons, Belgium.\\
E-mail: Garattini@mi.infn.it}
\title{Entropy and the cosmological constant: a spacetime-foam approach}

\begin{abstract}
A simple model of spacetime foam, made by $N$ wormholes in a semiclassical
approximation, is taken under examination. The Casimir-like energy of the
quantum fluctuation of such a model and its probability of being realized
are computed. Implications on the Bekenstein-Hawking entropy and the
cosmological constant are considered.
\end{abstract}

\maketitle

\section{\protect\bigskip Introduction}

The problem of merging General Relativity with Quantum Field Theory is known
as Quantum Gravity. One aspect of this merged theories is that at the Planck
scale, spacetime could be subjected to topology and metric fluctuations\cite
{Wheeler}. Such a fluctuating spacetime is known under the name of ``{\it %
spacetime foam}'' which can be taken as a model for the quantum
gravitational vacuum. At this scale of lengths (or energies) quantum
processes like black hole pair creation could become relevant. To establish
if a foamy spacetime could be considered as a candidate for a Quantum
Gravitational vacuum, we can examine the structure of the effective
potential for such a spacetime. It has been shown that flat space is the
classical minimum of the energy for General Relativity\cite{SY}. However
there are indications that flat space is not the true ground state when a
temperature is introduced, at least for the Schwarzschild space in absence
of matter fields\cite{GPY}. It is also argued that when gravity is coupled
to $N$ conformally invariant scalar fields the evidence that the
ground-state expectation value of the metric is flat space is false\cite{HH}%
. Moreover it is also believed that in a foamy spacetime, general relativity
can be renormalized when a density of virtual black holes is taken under
consideration coupled to $N$ fermion fields in a $1/N$ expansion\cite{CS}.
With these examples at hand, we have led to consider the possibility of
having a ground state different from flat space even at zero temperature.
Units in which $\hbar =c=k=1$ are used throughout the paper.

\section{Escaping from Flat Space}

If the effective potential (more precisely effective energy) is analyzed at
one loop in a Schwarzschild background, we discover that there exists an
imaginary contribution, namely flat space is unstable\cite{Remo1}. What is
the physical interpretation associated to this instability. We can begin by
observing that the ``simplest'' quantum process approximating a foamy
spacetime, in absence of matter fields, could be a black hole pair creation
of the neutral type. One possibility of describing such a process is
represented by the Schwarzschild-deSitter metric which asymptotically
approaches the deSitter metric. Its degenerate or extreme version is best
known as the Nariai metric\cite{Nariai}. Here we have an external
background, the cosmological constant $\Lambda _{c}$, which gives a nonzero
probability of having a neutral black hole pair produced with its components
accelerating away from each other. Nevertheless this process is believed to
be highly suppressed, at least for $\Lambda _{c}\gg 1$ in Planck's units\cite
{Bousso-Hawking}. In any case, metrics with a cosmological constant have
different boundary conditions compared to flat space. The Schwarzschild
metric is the only case available. Here the whole spacetime can be regarded
as a black hole-anti-black hole pair formed up by a black hole with positive
mass $M$ in the coordinate system of the observer and an {\it anti black-hole%
} with negative mass $-M$ in the system where the observer is not present.
In this way we have an energy preserving mechanism, because flat space has 
{\it zero energy} and the pair has zero energy too. However, in this case we
have not a cosmological {\it force} producing the pair: we have only pure
gravitational fluctuations. The black hole-anti-black hole pair has also a
relevant pictorial interpretation: the black hole with positive mass $M$ and
the {\it anti black-hole} with negative mass $-M$ can be considered the
components of a virtual dipole with zero total energy created by a large
quantum gravitational fluctuation\cite{Modanese}. Note that this is the only
physical process compatible with the energy conservation. The importance of
having the same energy behaviour ({\it asymptotic}) is related to the
possibility of having a spontaneous transition from one spacetime to another
one with the same boundary condition \cite{Witten}. This transition is a
decay from the false vacuum to the true one\cite{Coleman,Mazur}. However, if
we take account of a pair of neutral black holes living in different
universes, there is no decay and more important no temperature is involved
to change from flat to curved space. To see if this process is realizable we
need to compute quantum corrections to the energy stored in the boundaries.
These quantum corrections are pure gravitational vacuum excitations which
can be measured by the Casimir energy, formally defined as 
\begin{equation}
E_{Casimir}\left[ \partial {\cal M}\right] =E_{0}\left[ \partial {\cal M}%
\right] -E_{0}\left[ 0\right] ,  \label{i0}
\end{equation}
where $E_{0}$ is the zero-point energy and $\partial {\cal M}$ is a boundary.

\section{Building the foam}

\bigskip We begin to consider the following line element (Einstein-Rosen
bridge) related to a single wormhole $ds^{2}=$ 
\begin{equation}
-N^{2}\left( r\right) dt^{2}+\frac{dr^{2}}{1-\frac{2MG}{r}}+r^{2}\left(
d\theta ^{2}+\sin ^{2}\theta d\phi ^{2}\right)
\end{equation}
We\bigskip\ wish to compute the Casimir-like energy $\Delta E\left( M\right)
=E\left( M\right) -E\left( 0\right) $%
\begin{equation}
=\frac{\left\langle \Psi \left| H^{Schw.}-H^{Flat}\right| \Psi \right\rangle 
}{\left\langle \Psi |\Psi \right\rangle }+\frac{\left\langle \Psi \left|
H_{ql}\right| \Psi \right\rangle }{\left\langle \Psi |\Psi \right\rangle }
\end{equation}
by perturbing the three-dimensional spatial metric $g_{ij}=\tilde{g}%
_{ij}+h_{ij}$. $\Delta E\left( M\right) $ is computed in a WKB
approximation, by looking at the graviton sector (spin 2 or TT tensor) in a
Schr\"{o}dinger representation with trial wave functionals of the Gaussian
form by means of a variational approach. The Spin-two operator is defined as 
\begin{equation}
\left( \triangle _{2}\right) _{j}^{a}:=-\triangle \delta _{j}^{a}+2R_{j}^{a}
\end{equation}
where $\triangle $ is the Laplacian on a Schwarzschild background and $R_{j%
\text{ }}^{a}$ is the mixed Ricci tensor whose components are: 
\begin{equation}
R_{j}^{a}=diag\left\{ \frac{-2MG}{r_{{}}^{3}},\frac{MG}{r_{{}}^{3}},\frac{MG%
}{r_{{}}^{3}}\right\} .
\end{equation}
The total energy at one loop, i.e., the classical term plus the stable and
unstable modes respectively, is 
\[
\Delta E_{q.l.}+\Delta E_{s}+\Delta E_{u} 
\]
where $\Delta E_{q.l.}$ is the quasilocal energy. For symmetric boundary
conditions with respect to the bifurcation surface $S_{0}$ (such as this
case $E_{q.l.}=E_{+}-E_{-}=0$. When the boundaries go to spatial infinity $%
E_{\pm }=M_{ADM}$. The \noindent Stable modes contribution is 
\begin{equation}
\Delta E_{s}=-\frac{V}{32\pi ^{2}}\left( \frac{3MG}{r_{0}^{3}}\right)
^{2}\ln \left( \frac{r_{0}^{3}\Lambda ^{2}}{3MG}\right) .
\end{equation}
$\Lambda $ is a cut-off to keep under control the $UV$ divergence, we can
think that $\Lambda \leq m_{p}$. For the unstable sector, there is only {\bf %
one eigenvalue in S-wave.} This is in agreement with Coleman arguments on
quantum tunneling: the presence of a unique negative eigenvalue in the
second order perturbation is a signal of a passage from a false vacuum to a
true vacuum. The Rayleigh-Ritz method joined to a numerical integration
technique gives $E^{2}=-.\,\allowbreak 175\,41/\left( MG\right) ^{2},$ to be
compared with the value $E^{2}=-.\,\allowbreak 19/\left( MG\right) ^{2}$ of
Ref.\cite{GPY}.\bigskip\ How to eliminate the instability? We consider $%
N_{w} $\ coherent wormholes (i.e., non-interacting) in a semiclassical
approximation and assume that there exists a covering of $\Sigma $\ such
that $\Sigma =\cup _{i=1}^{N_{w}}\Sigma _{i}$, with $\Sigma _{i}\cap \Sigma
_{j}=\emptyset $\ when $i\neq j$. Each $\Sigma _{i}$\ has the topology $%
S^{2}\times R^{1}$\ with boundaries $\partial \Sigma _{i}^{\pm }$\ with
respect to each bifurcation surface. On each surface $\Sigma _{i}$,
quasilocal energy is zero because we assume that on each copy of the single
wormhole there is symmetry with respect to each bifurcation surface. Thus
the total energy for the collection is $E_{|2}^{tot}=N_{w}H_{|2}$ and the
total trial wave functional is the product of $N_{w}$ t.w.f. 
\begin{equation}
\Psi _{tot}^{\perp }=\Psi _{1}^{\perp }\otimes \Psi _{2}^{\perp }\otimes
\ldots \ldots \Psi _{N_{w}}^{\perp }
\end{equation}

By repeating the same calculations done for the single wormhole for the N$%
_{w}$\ wormhole system, we obtain

\begin{description}
\item[a)]  The total Casimir energy (stable modes), at its minimum, is 
\[
\Delta E_{s}\left( M\right) \sim -N_{w}^{2}\frac{V}{64\pi ^{2}}\frac{\Lambda
^{4}}{e}. 
\]
The minimum does not correspond to flat space $\rightarrow $ $\Delta
E_{s}\left( M\right) \neq 0.$

\item[b)]  The initial boundary located at $R_{\pm }$ will be reduced to $%
R_{\pm }/N_{w}.$

\item[c)]  Since the boundary is reduced there exists a critical radius $%
\rho _{c}=1.\,\allowbreak 113\,4$ such that : $\forall N\geq N_{w_{c}}\
\exists $ $r_{c}$ $s.t.$ $\forall \ r_{0}\leq r\leq r_{c},\ \sigma \left(
\Delta _{2}\right) =\emptyset $. This means that the system begins to be
stable\cite{Remo2,Remo3}. To be compared with the value $\rho _{c}=1.445$
obtained by B. Allen in Ref.\cite{Allen}.
\end{description}

\section{Area Spectrum, Entropy and the Cosmological constant\protect\bigskip%
}

Bekenstein has proposed that a black hole does have an entropy proportional
to the area of its horizon $S_{bh}=const\times A_{hor}$\cite{J.Bekenstein}.
In natural units one finds that the proportionality constant is set to $%
1/4G=1/4l_{p}^{2}$, so that the entropy becomes $S=A/4G=A/4l_{p}^{2}.$
Another proposal always made by Bekenstein is the quantization of the area
for nonextremal black holes $a_{n}=\alpha l_{p}^{2}\left( n+\eta \right) $%
\qquad $\eta >-1$\qquad $n=1,2,\ldots $ The area is measured by the quantity 
\begin{equation}
A\left( S_{0}\right) =\int_{S_{0}}d^{2}x\sqrt{\sigma }.
\end{equation}
We would like to evaluate the mean value of the area 
\begin{equation}
A\left( S_{0}\right) =\frac{\left\langle \Psi _{F}\left| \hat{A}\right| \Psi
_{F}\right\rangle }{\left\langle \Psi _{F}|\Psi _{F}\right\rangle }=\frac{%
\left\langle \Psi _{F}\left| \widehat{\int_{S_{0}}d^{2}x\sqrt{\sigma }}%
\right| \Psi _{F}\right\rangle }{\left\langle \Psi _{F}|\Psi
_{F}\right\rangle },
\end{equation}
computed on the foam state 
\begin{equation}
\left| \Psi _{F}\right\rangle =\Psi _{1}^{\perp }\otimes \Psi _{2}^{\perp
}\otimes \ldots \ldots \Psi _{N_{w}}^{\perp }.
\end{equation}
Consider $\sigma _{ab}=\bar{\sigma}_{ab}+\delta \sigma _{ab}\bar{\sigma}_{ab}
$ is such that $\int_{S_{0}}d^{2}x\sqrt{\bar{\sigma}}=4\pi \bar{r}^{2}$ and $%
\bar{r}$ is the radius of $S_{0}$%
\begin{equation}
A\left( S_{0}\right) =\frac{\left\langle \Psi _{F}\left| \hat{A}\right| \Psi
_{F}\right\rangle }{\left\langle \Psi _{F}|\Psi _{F}\right\rangle }=4\pi 
\bar{r}^{2}.
\end{equation}
Suppose to consider the mean value of the area $A$ computed on a given {\it %
macroscopic} fixed radius $R$. On the basis of our foam model, we obtain $%
A=\bigcup\limits_{i=1}^{N}A_{i}$, with $A_{i}\cap A_{j}=\emptyset $ when $%
i\neq j$. Thus 
\begin{equation}
A=4\pi R^{2}=\sum\limits_{i=1}^{N}A_{i}=\sum\limits_{i=1}^{N}4\pi \bar{r}%
_{i}^{2}.
\end{equation}
When $\bar{r}_{i}\rightarrow l_{p}$, $A_{i}\rightarrow A_{l_{P}}$ and\cite
{Remo3} 
\[
A=NA_{l_{P}}=N4\pi l_{p}^{2}\Longrightarrow S=\frac{A}{4l_{p}^{2}}=\frac{%
N4\pi l_{p}^{2}}{4l_{p}^{2}}=N\pi .
\]
Thus the macroscopic area is represented by $N$\ microscopic areas of the
Planckian size. In this sense we will claim that the area is quantized. The
first consequence is the mass quantization of the Schwarzschild black hole,
namely 
\begin{equation}
S=4\pi M^{2}G=4\pi M^{2}l_{p}^{2}=N\pi \Longrightarrow M=\frac{\sqrt{N}}{%
2l_{p}}.
\end{equation}
To be compared with Refs.\cite{Hod,Kastrup,Makela,Mazur1,Vaz}. A second
consequence is that in de Sitter space, the cosmological constant is
quantized in terms of $l_{p}$, i.e. 
\begin{equation}
S=\frac{3\pi }{l_{p}^{2}\Lambda _{c}}=\frac{A}{4l_{p}^{2}}=\frac{N4\pi
l_{p}^{2}}{4l_{p}^{2}}=N\pi \Longrightarrow \frac{3}{l_{p}^{2}N}=\Lambda
_{c}.
\end{equation}
It is possible to give an estimate of the total amount of Planckian
wormholes needed to fill the space beginning from the Planck era $\left(
\Lambda \sim \left( 10^{16}-10^{18}GeV\right) ^{2}\right) $ up to the space
in which we now live $\Lambda \leq \left( 10^{-42}GeV\right) ^{2}$. 
\begin{equation}
\frac{1}{N}10^{38}GeV^{2}=10^{-84}GeV^{2}\rightarrow N=10^{122},
\end{equation}
in agreement with the observational data $\Lambda _{c}\lesssim
10^{-122}l_{P}^{-2}$ coming from the Friedmann-Robertson-Walker cosmology
constraining the cosmological constant\cite{MVisser}.

\section{Problems and Conclusions}

There are several examples indicating that flat space cannot be taken as the
ground state of a quantum theory of gravity. In particular beginning with
the result that hot flat space is unstable with respect to a nucleation of a
single black hole, we have discovered that flat space is unstable with
respect to a neutral black hole pair creation. The instability generated by
the pair (wormhole) is stabilized if a certain large number of wormholes is
considered. Unfortunately, the model depends on a cutoff. Moreover,
contributions coming from the decompositions of the metric at one loop and
their interactions with the graviton sector are missing. Matter fields have
been completely neglected. Nevertheless there are good signals of a
spacetime foam: first of all the size of the energy fluctuations\cite
{Wheeler} $\Delta E\propto 1/L^{4}\propto \Lambda ^{4}.${\it \ }Area,
Entropy and the Cosmological constant are quantized as a simple consequence
of the covering property of our simple wormhole model of the foam.

\section{Acknowledgments}

I would like to thank the organizers and Prof. A. Perdichizzi for a partial
financial support. Special thanks to the organizers who have given to me the
opportunity of presenting this talk.

\end{document}